%% file: main.tex
\documentclass[sigconf, screen, authorversion]{acmart}  
\AtBeginDocument{%
  }

\copyrightyear{2026}
\acmYear{2026}
\setcopyright{rightsretained}  
\acmConference[SIGCSE TS 2026]{Proceedings of the 57th ACM Technical Symposium on Computer Science Education V.1}{February 18--21, 2026}{St. Louis, MO, USA}
\acmBooktitle{Proceedings of the 57th ACM Technical Symposium on Computer Science Education V.1 (SIGCSE TS 2026), February 18--21, 2026, St. Louis, MO, USA}
\acmDOI{10.1145/3770762.3772612}
\acmISBN{979-8-4007-2256-1/2026/02}

\include{macros}
\begin{document}


\title[Closing the Loop: An Instructor-in-the-Loop AI Assistance System]{Closing the Loop: An Instructor-in-the-Loop AI Assistance System for Supporting Student Help-Seeking in Programming Education}


\author{Tung Phung}
\affiliation{%
  \institution{MPI-SWS}
  \city{Saarbrücken}
  \country{Germany}}
\email{mphung@mpi-sws.org}

\author{Heeryung Choi}
\affiliation{%
  \institution{University of Minnesota}
  \city{Twin Cities}
  \country{USA}}
\email{heeryung@umn.edu}

\author{Mengyan	Wu}
\affiliation{%
  \institution{University of Michigan}
  \city{Ann Arbor}
  \country{USA}}
\email{mengyanw@umich.edu}

\author{Christopher	Brooks}
\affiliation{%
  \institution{University of Michigan}
  \city{Ann Arbor}
  \country{USA}}
\email{brooksch@umich.edu}

\author{Sumit Gulwani}
\affiliation{%
  \institution{Microsoft}
  \city{Redmond}
  \country{USA}}
\email{sumitg@microsoft.com}

\author{Adish Singla}
\affiliation{%
  \institution{MPI-SWS}
  \city{Saarbrücken}
  \country{Germany}}
\email{adishs@mpi-sws.org}

\renewcommand{\shortauthors}{Tung Phung et al.}

\input{0_abstract}

\ccsdesc[500]{Social and professional topics~Computing education}

\keywords{Programming Education, Feedback Generation, Human-AI Collaboration, Generative AI}

\maketitle

\input{1_introduction}
\input{2_related_work}

\input{3_tool_design}
\input{4_classroom_deployment}

\input{5_results_and_discussions}
\input{6_limitations}

\input{7_conclusion}

\begin{acks}
Funded/Co-funded by the European Union (ERC, TOPS, 101039090). Views and opinions expressed are however those of the author(s) only and do not necessarily reflect those of the European Union or the European Research Council. Neither the European Union nor the granting authority can be held responsible for them.
\end{acks}

\bibliographystyle{ACM-Reference-Format}
\balance
\bibliography{main}

\end{document}

%% file: macros.tex
\usepackage{multirow}
\usepackage{subcaption}
\usepackage{xspace}
\usepackage{colortbl}
\usepackage[inline]{enumitem}
\usepackage[dvipsnames]{xcolor}
\usepackage{soul}
\usepackage{makecell}
\usepackage{pifont}  
\usepackage{balance}

\definecolor{ExpHighlight}{rgb}{0.8, 0.1, 0.1}
\definecolor{mygreen}{rgb}{0,0.6,0}
\definecolor{mygray}{rgb}{0.5,0.5,0.5}
\definecolor{mymauve}{rgb}{0.58,0,0.82}
\definecolor{mynavy}{rgb}{0,0.133,0.302}
\definecolor{TextHighlight}{rgb}{1,1,0.6}
\definecolor{CodeGray}{rgb}{0.8,0.8,0.8}
\definecolor{CodeDarkGray}{rgb}{0.6,0.6,0.6}
\definecolor{OldPromptGrey}{rgb}{0.9,0.9,0.9}
\definecolor{NewPromptGreen}{rgb}{0.961,0.990,0.977}
\definecolor{InputOutput}{rgb}{0.764,0.345,0.009}
\definecolor{promptinputcolor}{rgb}{0.055,0.722,0.745}
\definecolor{promptheadercolor}{rgb}{0, 0, 0}

\setlist{nolistsep,leftmargin=*}

\newcommand{\graysquarequestion}{%
  \rlap{\textcolor{gray!70}{\rule{0.8em}{0.8em}}}%
  \textcolor{white}{\raisebox{0.1em}{\makebox[0.8em]{\textbf{\small ?}}}}%
}

%% file: 0_abstract.tex
\begin{abstract}

Timely and high-quality feedback is essential for effective learning in programming courses; yet, providing such support at scale remains a challenge. While AI-based systems offer scalable and immediate help, their responses can occasionally be inaccurate or insufficient. Human instructors, in contrast, may bring more valuable expertise but are limited in time and availability. To address these limitations, we present a hybrid help framework that integrates AI-generated hints with an escalation mechanism, allowing students to request feedback from instructors when AI support falls short. This design leverages the strengths of AI for scale and responsiveness while reserving instructor effort for moments of greatest need. We deployed this tool in a data science programming course with 82 students. We observe that out of the total 673 AI-generated hints, students rated 146 (22\%) as unhelpful. Among those, only 16 (11\%) of the cases were escalated to the instructors. A qualitative investigation of instructor responses showed that those feedback instances were incorrect or insufficient roughly half of the time. This finding suggests that when AI support fails, even instructors with expertise may need to pay greater attention to avoid making mistakes. We will publicly release the tool for broader adoption and enable further studies in other classrooms. Our work contributes a practical approach to scaling high-quality support and informs future efforts to effectively integrate AI and humans in education.

\end{abstract}

%% file: 1_introduction.tex


\section{Introduction}  \label{sec:introduction}

\input{illu_fig_main}

The growing capabilities of AI have created new opportunities to enhance computing education~\cite{DBLP:journals/corr/abs-2402-01580,DBLP:conf/icer/PhungPCGKMSS22,DBLP:conf/iticse/Prather00BACKKK23,franklin2025generative}, particularly in providing immediate and tailored feedback to students~\cite{gong2025impact,phung2025plan,liffiton2023codehelp}.
However, AI-generated feedback can lack the pedagogical quality and empathetic nuance provided by human instructors~\cite{phung2025bridging,er2025assessing,zhang2025evaluating}. Further, it cannot deliver \emph{instructor presence}---the sense that an instructor is actively engaged and accessible---which has been shown to significantly impact student engagement and learning experience~\cite{richardson2015conceptualizing,oyarzun2018instructor,ladyshewsky2013instructor}. 
Conversely, human instructors are limited by time and availability, making it challenging to offer support to all students~\cite{nguyen2024we,shein2019cs}. These complementary strengths and limitations motivate the need for hybrid approaches that can combine the advantages of both.

This paper describes the design and real-world deployment of a hybrid system that integrates AI and human instructors to collaboratively support students. Students can request on-demand hints from an embedded AI while working on assignments. If a student finds an AI hint unhelpful, they can choose to \emph{escalate} the issues to receive feedback from a human instructor (see Figure~\ref{fig:illustrative_example}). This escalation mechanism aims to serve both as a fallback for correctness and expertise, as well as a way to introduce empathy and a sense of instructor presence into the feedback process. Instead of positioning AI and instructors as separate sources of support, our system creates a coordinated workflow where instructors engage when most needed, maximizing both scalability and student experience.

We deployed this system in a real-world data science programming course. We aimed to understand how students and instructors utilize this hybrid model of support through the following RQs:

\begin{itemize} [leftmargin=0pt,itemsep=1pt,topsep=3pt,label={}]
    \item \textbf{RQ1}: How often do students request instructors versus AI?
    \item \textbf{RQ2}: How long does it take to receive instructor feedback for escalation requests and what do students do during that time?
    \item \textbf{RQ3}: Why do students choose to escalate to instructors and do instructors provide high-quality feedback in those cases?
\end{itemize}

Our contributions are threefold:

\begin{enumerate}
    \item We propose a hybrid system that allows students to request AI hints and escalate to instructors when needed, leveraging the complementary strengths of AI and human support.
    \item We report our experiences deploying this system in a real-world classroom setting, providing empirical insights into student behavior and learned lessons to better utilize the system.
    \item We will release our tool implementation to facilitate broader adoption and studies in diverse educational environments.
\end{enumerate}

Our work offers a pragmatic approach to scaling high-quality feedback in education and contributes to the broader discussion on how to effectively integrate AI into learning environments.

%% file: illu_fig_main.tex
\begin{figure*}[t]
        \scalebox{1}{
            \includegraphics[width=0.95\linewidth]{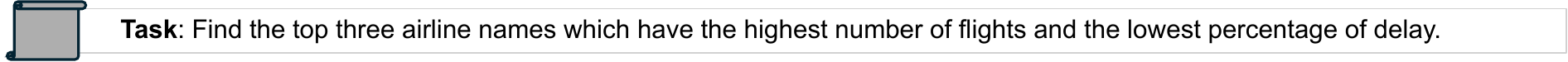}
        }
        \scalebox{1}{
            \includegraphics[width=0.95\linewidth]{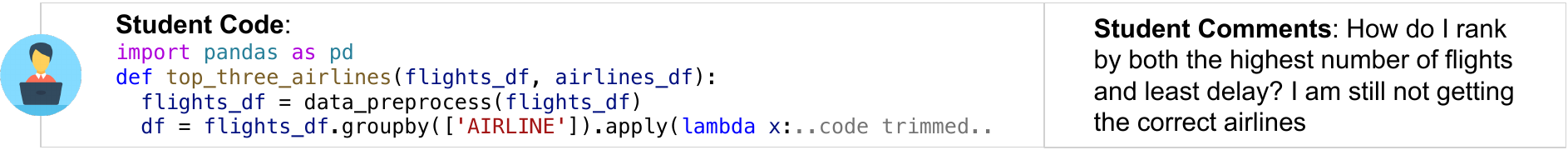}
        }
        \scalebox{1}{
            \includegraphics[width=0.95\linewidth]{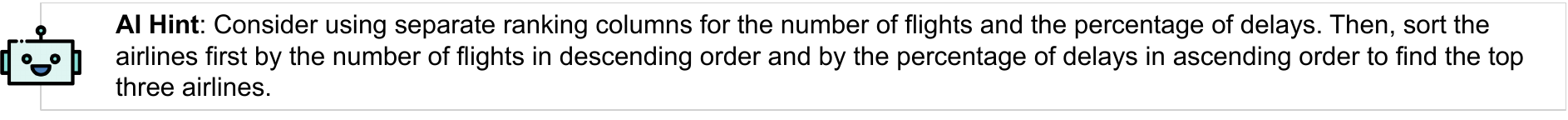}
        }
        \scalebox{1}{
            \includegraphics[width=0.95\linewidth]{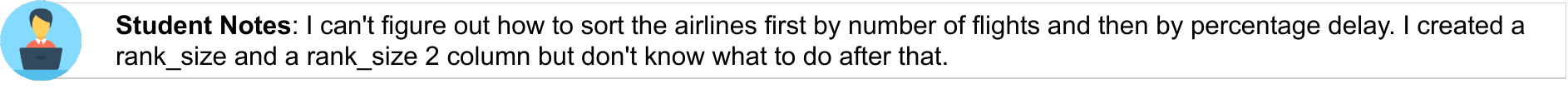}
        }
        \scalebox{1}{
            \includegraphics[width=0.95\linewidth]{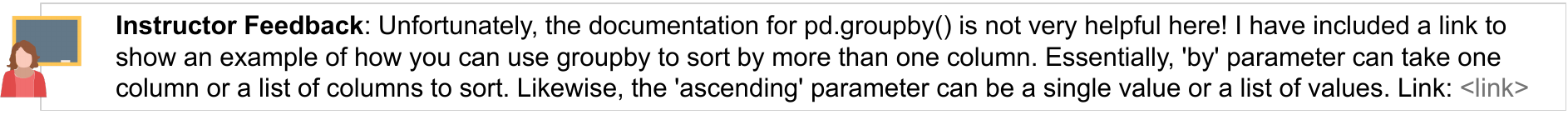}
        }
    \vspace{-3mm}
    \caption{
        Illustrative example. From top to bottom: Task description; Student's code and comments for requesting a debugging hint; AI-generated hint, which student then rated unhelpful; Student's notes for instructors when escalating for their feedback; and Instructor's feedback, which is more friendly, better targeted to student's issue, and contains reference to online resource.
    }
    \label{fig:illustrative_example}
    \vspace{-2mm}
\end{figure*}

%% file: 2_related_work.tex

\section{Related Work}  \label{sec:related_work}

\textbf{AI-generated feedback and hints.}
Recent work has increasingly explored AI for generating feedback and hints in programming education~\cite{DBLP:conf/icer/PhungPCGKMSS22,lohr2025you,DBLP:conf/sigcse/WoodrowMP24,DBLP:conf/nips/KotalwarGS24,kazemitabaar2024codeaid,fenu2024exploring}. Several studies have demonstrated the potential of AI to deliver support at scale. For instance, Ma et al.~\cite{DBLP:conf/aied/MaCK24} and Zhang et al.~\cite{DBLP:conf/aaai/ZhangD0PMX24} explored using AI-generated feedback in CS1 classrooms, reporting positive student perceptions. Xavier et al.~\cite{xavier2025empowering} found that AI support can substantially reduce instructor workload across educational contexts. 
Beyond generic feedback, some approaches have gone further by offering tailored assistance based on pedagogical theories. For example, Phung et al.~\cite{phung2025plan} implemented multiple AI-generated hint types aligned with metacognitive phases, finding these hints to be associated with higher performance. While these results highlight AI's promise for enhancing the learning experience at scale, there are still persistent limitations. Even advanced AI models can produce incorrect~\cite{DBLP:conf/sigcse/AhmedSLK25}, misleading~\cite{DBLP:conf/ace/RoestKJ24}, or inconsistent~\cite{DBLP:conf/uemcom/ZabalaN24} feedback, and their quality often remains below that of human instructors~\cite{DBLP:conf/icer/PhungPCGKMSS22,er2025assessing}. Moreover, students may be skeptical of feedback generated by AI regardless of its quality. Nazaretsky et al.~\cite{nazaretsky2024ai} found that the source of the feedback (i.e., AI or humans) alone affects students' assessment: they favor feedback that they know comes from humans and decrease evaluation if the feedback is provided by AI. These challenges necessitate human involvement in the feedback process to complement AI and build student trust. Our work addresses this gap by introducing human instructors in the loop, allowing students to escalate to instructors when AI hints are insufficient.

\input{student_flow_fig_main}

\textbf{Human-AI collaboration for providing feedback.}
Recent research has explored several models of human-AI collaboration for providing feedback~\cite{pahi2024enhancing,DBLP:conf/sigcse/AhmedSLK25,zhang2025evaluating}. Ahmed et al.~\cite{DBLP:conf/sigcse/AhmedSLK25} studied a pipeline in which AI drafts feedback that is then reviewed and possibly adjusted by teaching assistants (TAs). However, this method did not translate to better student performance, and the authors noted that some TAs became overly reliant on AI drafts, resulting in failure to identify and correct inaccuracies. In contrast, in our approach, human instructors know the AI-generated hints were unhelpful and thus, naturally avoid this complacency issue.
Zhang et al.~\cite{zhang2025evaluating} applied a different model, leveraging AI to revise and improve feedback drafted by humans. Their study in a university-level programming course showed that students perceived human-written and human-AI co-produced feedback as more credible than feedback generated solely by AI. However, this approach requires human instructors to produce feedback for all help requests, limiting its scalability and potentially delaying response time. 
In comparison, our system employs AI to handle the majority of the workload, with human instructors only stepping in when necessary, thus ensuring timely feedback for most requests while still keeping humans in the loop.

%% file: student_flow_fig_main.tex
\begin{figure*}[t]
    \centering
    \scalebox{1}{
        \includegraphics[width=\linewidth]{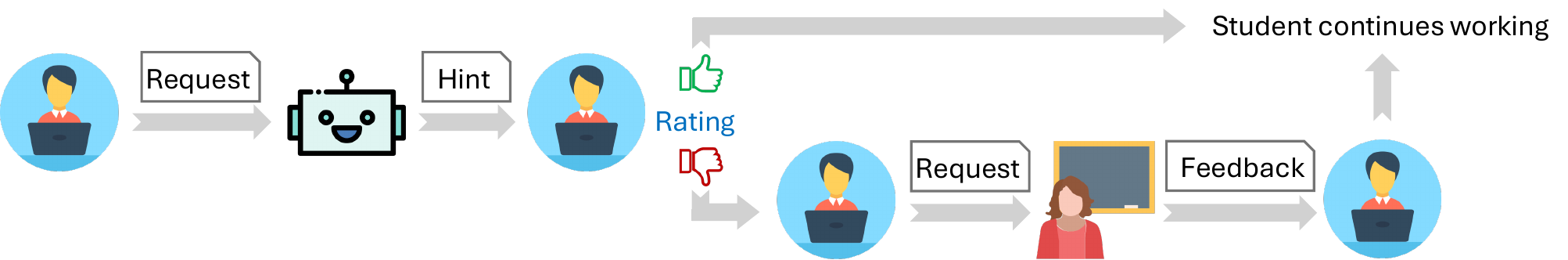}
    }
    \vspace{-6mm}
    \caption{
        Illustration of student requesting help. If AI hint is unhelpful, students can choose to escalate to human instructors.
    }
    \label{fig:student_flow}
    \vspace{-0.5mm}
\end{figure*}

%% file: 3_tool_design.tex

\input{student_interface_fig_main}

\section{Tool Design}  \label{sec:tool}

We design a help system that enables students to receive tailored support from both AI and human instructors directly within their programming environment. The system allows students to request AI-generated hints on demand and, especially, offers a mechanism for students to escalate to instructors when AI support is perceived as unhelpful (see Figure~\ref{fig:student_flow}). The tool comprises three modular components: (1) a backend for generating AI-powered hints, (2) a student-facing interface for requesting and receiving help, and (3) an instructor-facing interface for providing feedback.\footnote{\url{https://github.com/machine-teaching-group/sigcse2026-closing-the-loop}} Each component is described in detail below.

\subsection{AI Hint Generation Backend}  \label{sec:tool.backend}
Inspired by prior work on scaffolding in programming education, the AI backend supports three types of hints: planning, debugging, and optimization~\cite{phung2025plan,rum2017metocognitive,alhazmi2020impact}. Planning hints guide students in outlining a solution strategy, debugging hints assist in identifying and fixing errors, and optimization hints help with improving code performance and readability. The techniques to generate these hints are adapted from literature~\cite{phung2025plan,rum2017metocognitive,alhazmi2020impact}. For debugging hints, the backend runs student's code to obtain the buggy output, prompts the AI to generate a corrected program, and then uses all this information, along with any student's comments (detailed in Section~\ref{sec:tool.student_interface}) to query AI for a hint about one bug in the code. Planning and optimization hints are generated using similar pipelines but with prompts adjusted to focus on solving strategy and code improvement, respectively. All hints were generated using GPT-4o~\cite{hurst2024gpt}.

The backend also manages communication logistics: It notifies instructors of new escalation requests and informs students when instructor feedback is available via email.

\subsection{Student Interaction Interface}  \label{sec:tool.student_interface}

Figure~\ref{fig:student_interface} illustrates the student interaction interface. To allow seamless access during coding, the interface is integrated directly into the student's programming environment, which in our case is JupyterLab (see Section~\ref{sec:classroom_deployment}). For each programming question, students can request AI hints by clicking on specific buttons for the hint types. To prevent over-reliance on hints, we impose per-question quotas: one planning hint, three debugging hints, and one optimization hint. When a student clicks a hint button, a dialog appears, allowing them to optionally write brief comments on their current progress or issues. These comments serve two purposes: (1) encouraging metacognition by prompting students to articulate their difficulties, and (2) providing AI with context to generate more relevant hints. The dialog also informs students that AI hints may take up to two minutes to generate, setting expectations for the service. After receiving a hint, students can rate it as \emph{Helpful} or \emph{Unhelpful}. If marked unhelpful, students are given the option to escalate to instructors for further feedback, noting that it might take instructors up to 24 hours to respond. For the escalation request, students can add textual notes to specify their doubts or clarify why they found the AI hint unhelpful. They are also informed that escalation requests are anonymous to the instructors. This design is aimed at reducing students' fear of being judged, offering a more approachable alternative to traditional support channels such as email.

\input{instructor_interface_fig_main}

All AI hints and instructor feedback are displayed in collapsible widgets below the hint buttons, ensuring easy access without clustering the interface. They are persisted across sessions and remain accessible even after closing and reopening notebooks.

\subsection{Instructor Feedback Interface}  \label{sec:tool.instructor_interface}

Figure~\ref{fig:instructor_interface} shows the web-based interface for instructors, designed to streamline the process of reviewing student requests and responding with feedback. Upon accessing, instructors are presented with the oldest unresolved request, including relevant information, such as the student's notebook, the student's comments, the AI-generated hint, and any notes provided. Instructors can then compose a targeted feedback message to respond to the student.

%% file: student_interface_fig_main.tex
\begin{figure}[t]
    \centering
    \scalebox{1}{
        \includegraphics[width=\linewidth]{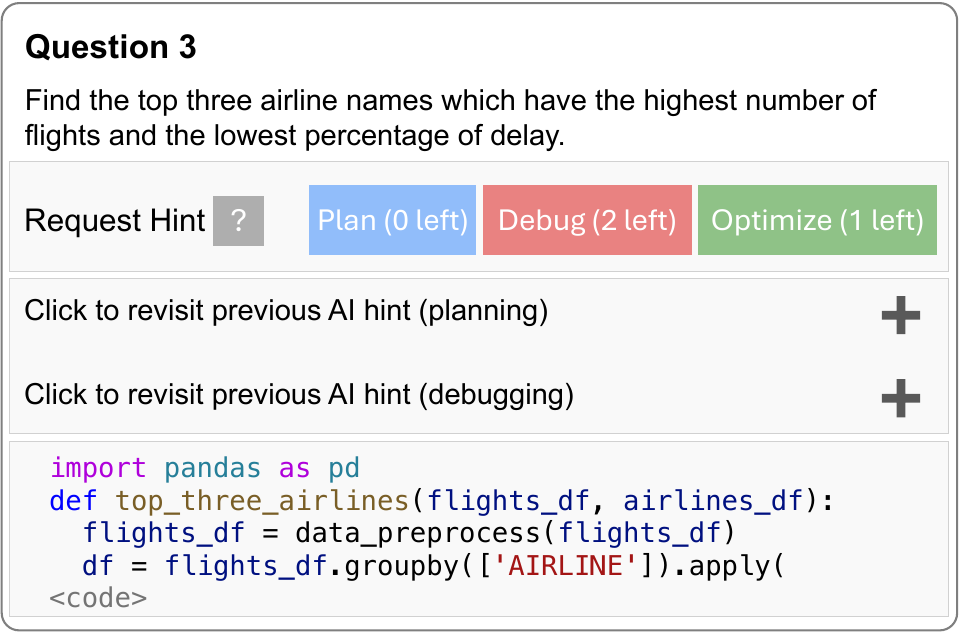}
    }
    \vspace{-5.5mm}
    \caption{
        Student interface for requesting AI hints. They can click the \graysquarequestion{} button to view a description of hint types, the three hint buttons for requesting AI hints, and the collapsible widgets to view received AI hints and instructor feedback.
    }
    \label{fig:student_interface}
    \vspace{-1.5mm}
\end{figure}

%% file: instructor_interface_fig_main.tex
\begin{figure}[t]
    \centering
    \scalebox{1}{
        \includegraphics[width=\linewidth]{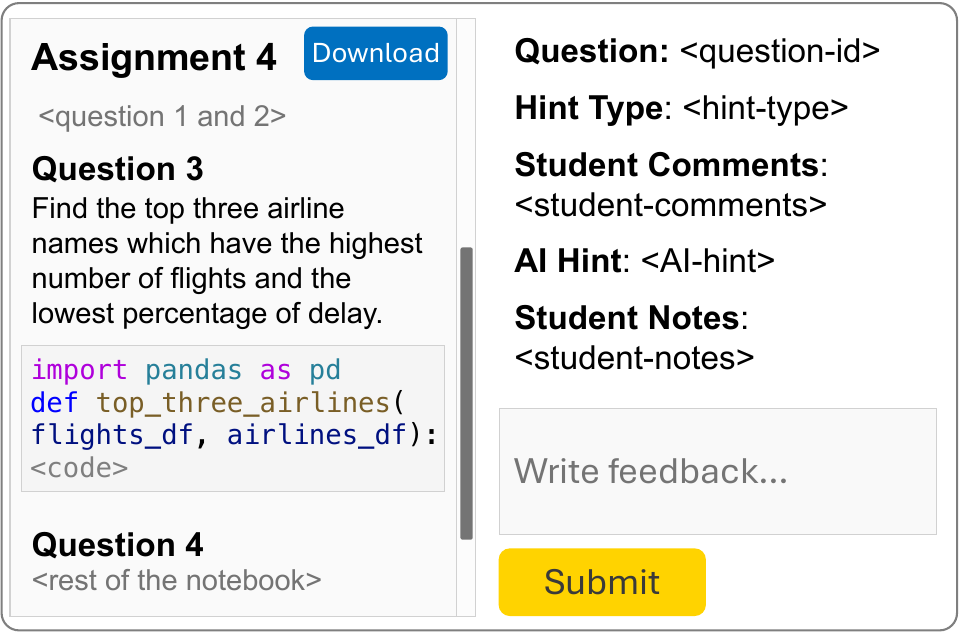}
    }
    \vspace{-5.5mm}
    \caption{
        Instructor interface for giving feedback as a single-page website. On the left is the student's notebook. On the right is other contextual information and a place for the instructor to input their feedback.
    }
    \label{fig:instructor_interface}
    \vspace{-1.5mm}
\end{figure}

%% file: 4_classroom_deployment.tex

\section{Classroom Deployment}  \label{sec:classroom_deployment}

\textbf{Course context.} 
We deployed this system in a credit-bearing introductory course on Python programming for data science, offered online as part of a Master's degree at the University of Michigan. This four-week course had weekly modules of increasing complexity, covering topics such as regular expressions, data frame manipulation, and handling CSV and Excel files. Each module included video lectures and an assignment in the form of a Jupyter notebook, containing three to four programming questions (14 questions in total). Overall, 82 students enrolled in the course. Students could submit each assignment multiple times, with their highest score used for grading, resulting in most achieving full marks for all assignments.
The task of responding to students' escalation requests was assigned to one instructor on the three-person instructional team as part of their regular instructional duties.

\textbf{Consent and data collection.}
Students were informed about the use of third-party services (OpenAI) for hint generation and the possibility that AI-generated hints may not always be correct. Upon requesting the first hint, students were presented with a consent pop-up showing all these details, and they could only request hints after agreeing to this notice. Requesting help was optional, with no other incentives or penalties. Students were also assured that all help requests were anonymous to the instructors. This study was reviewed and classified as exempt from oversight by the Institutional Review Board under application number HUM00251143. During the course, we collected interaction data, including records of help requests, ratings of AI hints, assignment submissions, and learning behaviors such as coding and watching lecture videos.

%% file: 5_results_and_discussions.tex

\section{Results and Discussion}  \label{sec:results}  

This section presents the outcomes from our classroom deployment.

\subsection{RQ1: Frequency of Escalation Requests}  \label{sec:results.rq1}

\subsubsection{Analysis Setup.} 
We examine the frequency of escalation requests and compare them with the use of AI-generated hints. Our analysis focuses on two aspects: (1) the number of requests by hint type, and (2) the distribution of requests across assignments.

\subsubsection{Findings and discussions.} 
Figure~\ref{fig:rq1} presents the results for RQ1. The high adoption rate of AI hints (87\% of students; 673 requests) consolidates prior findings that students are generally receptive to intelligent tutoring systems~\cite{DBLP:conf/aied/MaCK24,stohr2024perceptions}. Out of the generated hints, 146 (22\%) were rated as unhelpful, and nine students escalated a total of 16 of those requests. This indicates that AI hints were often deemed useful by students, but there were still cases where additional support was needed.
Among three hint types, debugging hints had the highest rate of unhelpful ratings (26\% rated unhelpful) and were the only type that led to escalation. This result aligns with previous findings on the limitations of automated systems in handling complex, context-specific debugging issues~\cite{phung2025bridging,jacobs2024evaluating}, and underscores the importance of human oversight to offer a pathway for richer pedagogical aid when AI falls short.

\input{RQ1_fig_main}

\input{RQ2_fig_main}

Figure~\ref{fig:rq1.count_by_assignment} shows that both types of help requests, AI hints and instructor feedback, were spread across all assignments. Assignments 3 and 4 had the highest number of requests, likely due to their increased complexity. While students requested the most AI hints for Assignment 4, they required instructor feedback most frequently for Assignment 3. This suggests that the need for AI hints and instructor feedback may not always be correlated: students might require different types of help in different situations, and thus, it is important to provide both AI hints and instructor feedback to meet diverse student needs.
Although later assignments had more requests overall, earlier assignments also saw substantial use, indicating that students were able to quickly become familiar with and make use of the system when needed.

\subsection{RQ2: Students' Waiting Time for Instructors}    \label{sec:results.rq2}
\subsubsection{Analysis setup.}
To address this RQ, we analyze the time students waited for instructor feedback, along with their learning behaviors during this period. Specifically, we examine whether students continued coding, watched lecture videos, requested additional AI hints, and whether they successfully solved the question.

\subsubsection{Findings and discussions.} As shown in Figure~\ref{fig:rq2}, the waiting time for instructor feedback varied, with an average of 13.5 hours. Most of this delay occurred while the instructor was offline. Once a request was viewed by the instructor, the average response time was 17.8 minutes. In contrast, AI-generated hints were delivered in approximately 20 seconds. This difference in waiting time likely contributed to the relatively low number of escalation requests, consistent with prior work reporting that most students canceled their help request after waiting for less than an hour~\cite{gao2023too}.
Addressing these delays could support more effective integration of human instructors within the help system. Potential solutions include distributing the workload across instructors in different time zones to improve coverage in global online courses~\cite{DBLP:conf/iticse/MalikWWP24}, providing students with an indicator showing if instructors are online and available for providing feedback to set clearer expectations about response times, or leveraging students who have finished the assignment as peer helpers in the escalation process~\cite{DBLP:conf/sigcse/MalikWP24,DBLP:conf/lak/Singh0WLKW24}.

Despite the delays, students remained engaged during this period (see Figure~\ref{fig:rq2}). Students always continued coding, with 87.5\% continuing to code within the first hour after escalating a question, and they also consulted lecture videos (75\% of the time). These patterns suggest a positive learning behavior, where the system did not render students overly reliant on feedback, but instead continued to engage with course materials and improve their solutions. 
During their attempts, students sought additional help, as shown by further AI hints being requested roughly half of the time. However, students were able to solve the question before receiving instructor help in only 25\% of the time, emphasizing the importance of instructor feedback in such difficult moments.

\input{RQ3a_fig_main}

\input{RQ3b_fig_main}

\subsection{RQ3: Escalation Reasons and Feedback}    \label{sec:results.rq3}

\subsubsection{Analysis setup.} 
To understand why students chose to escalate to instructors, we focus on the nine students who used the escalation feature and compare cases when they escalated versus when they did not. 
One author of this paper, who has extensive experience in teaching Python, carefully analyzed students' code, comments, AI hints, and escalation notes to identify the bug types in their code and the reasons for rating AI hints as unhelpful. 
Bug types are classified using a taxonomy of mistakes in data science programming from the literature~\cite{singh2024investigating}, including: (1) Dataset misunderstanding (misreading the dataset or its schema), (2) Task misunderstanding (misinterpreting requirements), (3) Mishandling of missing values, (4) Semantic bugs (incorrectly using a library, function, or operator), (5) Language or environment bugs, and (6) Suboptimal coding (working but non-optimal code). A single code may contain multiple bug types.
To categorize the reasons students rated AI hints as unhelpful, we adopt a coding scheme based on prior work~\cite{phung2025bridging}, which includes: (1) Incorrect (hint was partly or fully wrong), (2) Uninformative (student wanted more information or guidance), (3) Misfocused (hint did not address the student's main issue), and (4) Unclear (student could not understand the hint).

Finally, instructor feedback is assessed using a similar procedure and the same coding scheme used for AI hints: feedback is regarded as high-quality if it meets all quality criteria; otherwise, it is low-quality for being incorrect, uninformative, misfocused, or unclear.

\subsubsection{Findings and discussions.}
As shown in Figure~\ref{fig:rq3.bug_types}, different bug types show similar escalation rates. When considering the qualities of the AI hints, however, escalation due to incorrectness (first bar in Figure~\ref{fig:rq3.reasons_unhelpful}) stands out, with 54\% of such cases resulting in escalation to instructors. This result may suggest a pattern in students' prediction of AI quality: when AI is incorrect, students may anticipate that the issue is difficult for AI to resolve and it will not be able to help even with retries, leading them to escalate to seek instructors' support. This finding further emphasizes the importance of humans in the system to intervene when AI hallucinates and generates misleading, unreasonable outputs~\cite{janeafik2024problem,kamel2024understanding}.

Figure~\ref{fig:rq3b} illustrates instructor feedback quality and how it varies by bug types and reasons AI hints were unhelpful.
Notably, instructor feedback had high quality in only 44\% of cases (seven out of 16). Furthermore, low-quality feedback was especially prevalent when AI-generated hints had already been incorrect. Among the seven escalations following incorrect AI hints, six (86\%) received instructor feedback that is classified as low-quality. 
Contrary to the common assumption that instructors can reliably deliver high-quality help~\cite{er2025assessing,yang2021power}, our results underscore that instructors, too, can struggle, particularly in complex cases when AI has already failed. This observation also points to opportunities for improvement, such as tools to assist instructors in diagnosing student mistakes and clarifying student doubts, workflows that encourage correcting student code before offering feedback, or simply reminding instructors to pay closer attention in those tricky situations.

\subsection{Student and Instructor Perspectives}    \label{sec:results.perspectives}

\textbf{Students' perspectives.}
There was an optional, anonymous survey administered at the end of the course to ask for students' opinions about the course content, instructors, and the help system. Sixteen students (20\%) responded. While we did not receive complaints or suggestions for improving the system, some students offered positive comments, such as "\textit{hint system was super helpful}".

\textbf{Instructors' perspectives.} The instructor who handled escalation requests also shared reflections on the system's design, noting that "\textit{the information provided in the initial request was really helpful}" and that "\textit{it was an interesting way to engage with the material and the students}". 
They also pointed to improvements, such as adding a feature to enable continued conversation with students: "\textit{I couldn't tell if there was a way to continue chatting through that application}" and another feature to let them know how the students evaluate their feedback: "\textit{I don't know if there's any further feedback from the student on whether my hint was helpful}". Overall, the instructional team of the course expressed pleasure with the system and interest in using it in future course iterations. 

%% file: RQ1_fig_main.tex
\begin{figure}
    \begin{minipage}{\linewidth}
        \centering
        \scalebox{1}{
            \includegraphics[height=0.03\paperheight]                {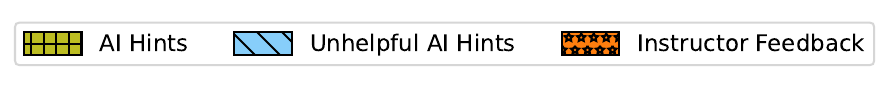}
        }
        \vspace{-5.5mm}
    \end{minipage}
    \begin{minipage}{0.45\linewidth}
        \centering
        \scalebox{1}{
            \includegraphics[height=0.15\paperheight]                {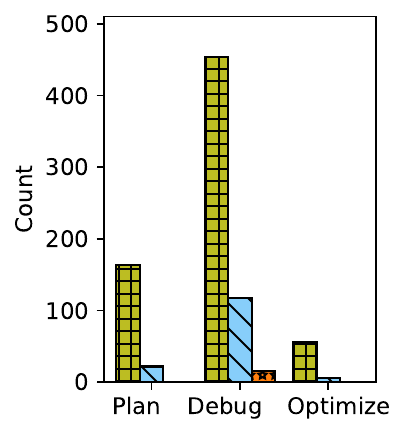}
        }
        \vspace{-6.5mm}
        \subcaption{Help requests by type}
        \label{fig:rq1.count_by_type}
    \end{minipage}
    \hfill
    %
    \begin{minipage}{0.53\linewidth}
        \centering
        \scalebox{1}{
            \includegraphics[height=0.15\paperheight]{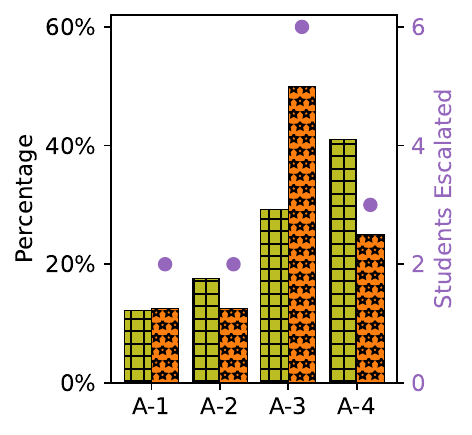}
        }
        \vspace{-6.5mm}
        \subcaption{Help requests by assignment}
        \label{fig:rq1.count_by_assignment}
    \end{minipage}
    \vspace{-3mm}
    \caption{
        Frequency of help requests. In total, 71 students (87\%) requested AI hints and nine (11\%) escalated to instructors. 
        (a) compares help requests by hint types, with only debugging hints being chosen for escalation (16 escalations in total).
        (b) compares help requests by assignments. Bar heights (left axis) indicate the percentage of total requests each assignment received. For instance, the first assignment accounted for about 15\% of all hint requests, and also roughly 15\% of the total instructor escalations. The purple dots (right axis) show the number of unique students who escalated. While help requests increased for later assignments, early assignments also observed substantial numbers of requests.
    }
    \vspace{-3mm}
    \label{fig:rq1}
\end{figure}

%% file: RQ2_fig_main.tex
\begin{figure*}[t]
    \centering
    \begin{minipage}{\linewidth}
    \centering
        \includegraphics[height=0.18\paperheight]{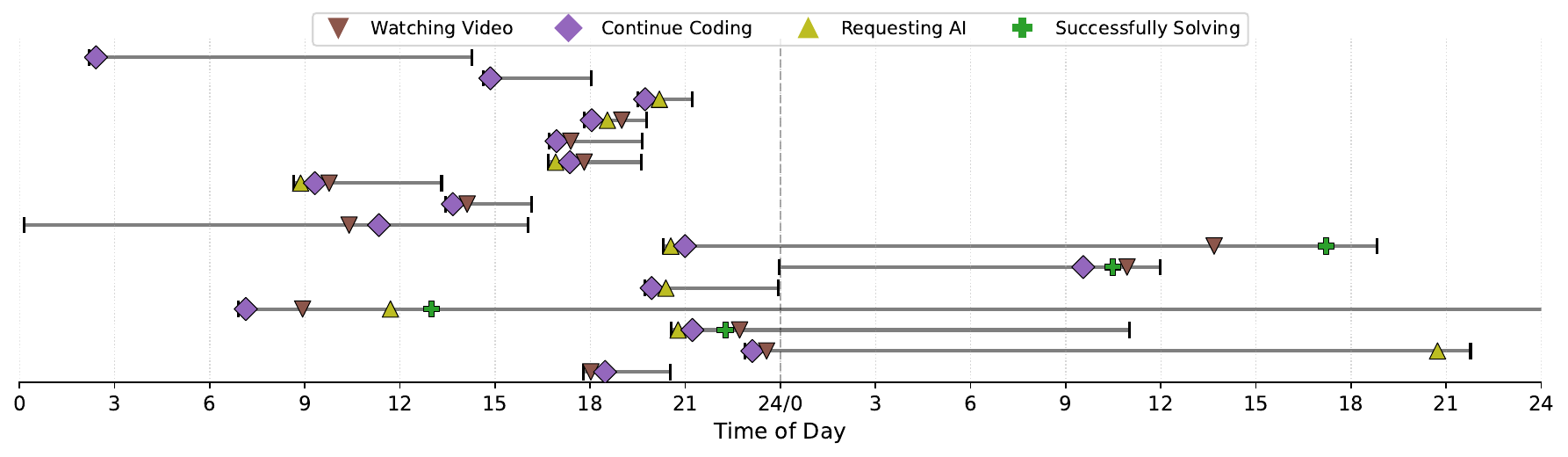}
    \end{minipage}
    \vspace{-5mm}
    \caption{Students' actions and progress while awaiting instructor feedback for all 16 escalations. Each horizontal line represents an escalation: From the \emph{time of day} when the student escalated to when instructor feedback was available, usually in the same or the next day---the fourth line from the bottom shows the only case where feedback took more than a day, for which we trimmed the line at the end of the next day. The markers denote the \emph{first} occurrences of activity or progress types:  $\textcolor{Brown}{\blacktriangledown}$ for watching lecture videos, \textcolor{Orchid}{\rotatebox[origin=c]{45}{$\blacksquare$}} for coding, $\textcolor{GreenYellow}{\blacktriangle}$ for requesting AI hints, and \textcolor{ForestGreen}{\ding{58}} for successfully solving the question. Students remained actively engaged during the wait, showing no sign of over-reliance on instructor feedback. Still, only 25\% of these cases were solved before receiving feedback, showing the difficulty of these instances and the importance of having instructor support.
    }
    \label{fig:rq2}
    \vspace{-2.5mm}
\end{figure*}

%% file: RQ3a_fig_main.tex
\begin{figure*}
    \centering
    \begin{subfigure}{\linewidth}
        \centering
        \scalebox{1}{
            \includegraphics[height=0.03\paperheight]{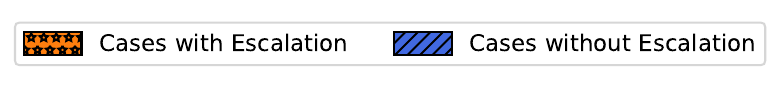}
        }
        \vspace{-2.5mm}
    \end{subfigure}
    \begin{subfigure}{0.55\linewidth}
        \centering
        \scalebox{1}{
            \includegraphics[height=0.09\paperheight]{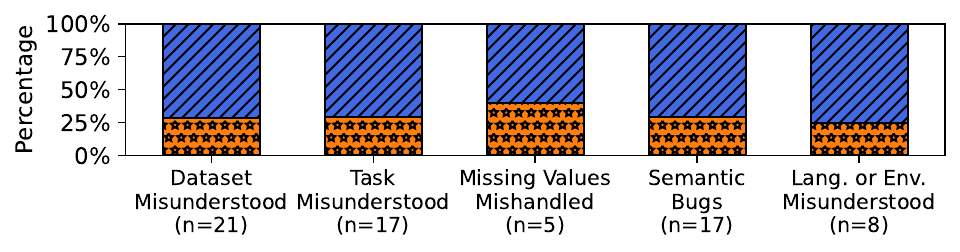}
        }
        \vspace{-6mm}
        \subcaption{Cases grouped by bug types}
        \label{fig:rq3.bug_types}
    \end{subfigure}
    \hfill
    \begin{subfigure}{0.445\linewidth}
        \centering
        \scalebox{1}{
            \includegraphics[height=0.09\paperheight]{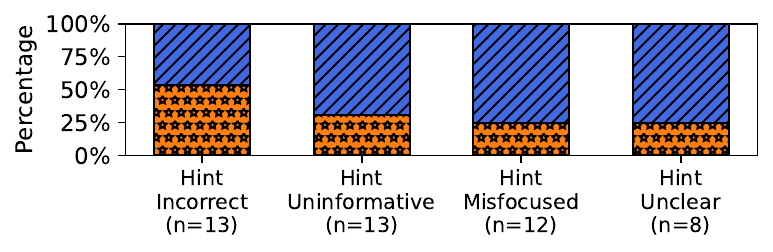}
        }
        \vspace{-6mm}
        \subcaption{Cases grouped by reasons causing AI hints unhelpful}
        \label{fig:rq3.reasons_unhelpful}
    \end{subfigure}
    \vspace{-7.5mm}
    \caption{
        Comparison of cases when students escalated versus when they did not. Included are all requests from the nine students who escalated at least once, totalling 46 instances: 16 with escalations and 30 without escalations. (a) compares the cases by bug types (one code may contain multiple bugs); the Suboptimal coding type is omitted due to zero counts. (b) compares the cases by reasons causing AI hints to be rated as unhelpful, with Hint Incorrect exhibiting the highest rate of escalation.
    }
    \label{fig:rq3}
    \vspace{-2mm}
\end{figure*}

%% file: RQ3b_fig_main.tex
\begin{figure*}
    \centering
    \begin{subfigure}{\linewidth}
        \centering
        \scalebox{1}{
            \includegraphics[height=0.03\paperheight]{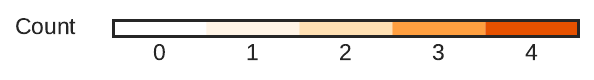}
        }
        \vspace{-2mm}
    \end{subfigure}
    \begin{subfigure}{0.475\linewidth}
        \centering
        \scalebox{1}{
            \includegraphics[height=0.09\paperheight]{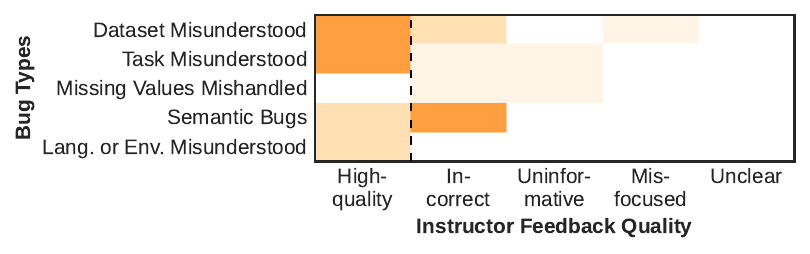}
        }
        \vspace{-2mm}
        \subcaption{Instructor feedback quality by bug types}
        \label{fig:rq3b.bug_types}
    \end{subfigure}
    \hfill
    \begin{subfigure}{0.5\linewidth}
        \centering
        \scalebox{1}{
            \includegraphics[height=0.09\paperheight]{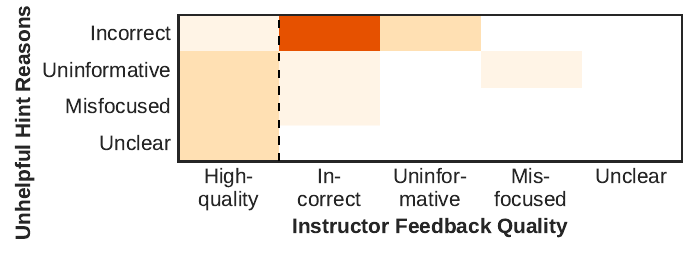}
        }
        \vspace{-2mm}
        \subcaption{Instructor feedback quality by reasons causing AI hints unhelpful}
        \label{fig:rq3b.reasons_unhelpful}
    \end{subfigure}
    \vspace{-3.5mm}
    \caption{
        Instructor feedback quality when AI hints were unhelpful and students escalated their help requests. (a) shows feedback quality by bug types, and (b) shows feedback quality by the reasons causing AI hints to be unhelpful. Approximately half of the feedback instances did not meet all quality criteria, with the majority occurring when AI hints were incorrect. This suggests the challenging nature of such cases, requiring careful instructor attention to provide effective support.
    }
    \label{fig:rq3b}
    \vspace{-3.5mm}
\end{figure*}

%% file: 6_limitations.tex

\subsection{Limitations}  \label{sec:limitations}
Our deployment was limited to a single course focused on introductory data science programming with a relatively small number of escalation requests, and this naturally limits the generalizability of our findings. Future research should integrate the system into a wider range of courses and contexts to validate and extend these results. In addition, our study did not measure the educational impact of the help system on long-term student performance, which should be assessed in future studies. The AI support in our system is currently designed as button-based, single-shot hints. Exploring alternative interfaces, such as chat-based AI tutors, could result in additional, complementary experience on how having instructors in the loop could benefit in various settings. Finally, expanding the system to include additional forms of assistance, such as peer feedback, could provide valuable insights into how different supports influence student learning behaviors and outcomes.

%% file: 7_conclusion.tex

\section{Conclusions}  \label{sec:conclusion}
This paper presents our experience deploying a hybrid help system that combines AI-generated hints with an instructor escalation mechanism to support students in programming education.
Findings from a real-world classroom deployment demonstrate the potential of the system and offer actionable insights for its effective use. While this approach can leverage the advantages of both AI (availability and scalability) and instructors (expertise and reaffirming instructor presence), ensuring quality remains essential and requires careful consideration.
More broadly, our work underscores the promise of hybrid systems in optimizing for both scale and quality in education, offering a practical model for integrating human and AI support to enhance learning environments.